\documentclass[aps,pra,superscriptaddress,amsmath,amssymb,preprintnumbers,twocolumn,showpacs]{revtex4}
\usepackage{amssymb}
\usepackage{graphicx}
\usepackage{bm}
\usepackage{color}

\newcommand{\Ignore}[1]{}

\newcommand{\Bra}[1]{\left\langle#1\right\vert}
\newcommand{\Ket}[1]{\left\vert#1\right\rangle}
\newcommand{\BraKet}[2]{\left\langle#1\vert #2\right\rangle}
\newcommand{\KetBra}[2]{\left\vert#1\right\rangle\left\langle#2\right\vert}
\newcommand{\Projector}[1]{\KetBra{#1}{#1}}

\newcommand{\KetBraSub}[3]{\Ket{#1}_{#3}\hspace*{-0.2mm}\Bra{#2}}

\newcommand{\trace}{\mathrm{tr}}

\newcommand{\purity}{\mathrm{P}}

\newcommand{\totsys}{\text{\rm tot}}
\newcommand{\mstsys}{\text{\rm X}}
\newcommand{\slvsys}{\text{\rm S}}

\newcommand{\up}{\uparrow}
\newcommand{\dn}{\downarrow}
\newcommand{\probup}{p_{\up\up}}
\newcommand{\absc}{\tilde{c}}

\newcommand{\Sig}[2]{\hat{\sigma}_{#1}^{(#2)}}

\begin{document}

% --------------------  TITLE  --------------------

\title{Oscillations of the purity in the repeated-measurement-based generation of quantum states}

% ------------  AUTHORS AND AFFILIATIONS ----------

\author{B. Militello}
\email{bdmilite@fisica.unipa.it} \affiliation{MIUR and
Dipartimento di Scienze Fisiche ed Astronomiche
dell'Universit\`{a} di Palermo, Via Archirafi 36, I-90123 Palermo,
Italy}
\author{K. Yuasa}
\affiliation{Waseda Institute for Advanced Study, Waseda
University, Tokyo 169-8050, Japan} \affiliation{Department of
Physics, Waseda University, Tokyo 169-8555, Japan}
\author{H. Nakazato}
\affiliation{Department of Physics, Waseda University, Tokyo
169-8555, Japan}
\author{A. Messina}
\affiliation{MIUR and Dipartimento di Scienze Fisiche ed
Astronomiche dell'Universit\`{a} di Palermo, Via Archirafi 36,
I-90123 Palermo, Italy}

\date{\today}

% --------------------  ABSTRACT  --------------------

\begin{abstract}
Repeated observations of a quantum system interacting with another
one can drive the latter toward a particular quantum state,
irrespectively of its initial condition, because of an {\em
effective non-unitary evolution}. If the target state is a pure
one, the degree of purity of the system approaches unity, even
when the initial condition of the system is a mixed state. In this
paper we study the behavior of the purity from the initial value
to the final one, that is unity. Depending on the parameters,
after a finite number of measurements, the purity exhibits
oscillations, that brings about a lower purity than that of the
initial state, which is a point to be taken care of in concrete
applications.
\end{abstract}


\maketitle

% --------------------  INTRODUCTION  ------------------

\section{Introduction}\label{sec:introduction}

The initialization of a physical system into a prefixed quantum
state is a fundamental task in connection with applications in
nano-technology and quantum information~\cite{ref:q_tech}. In
particular, the purification of a quantum state is an important
issue in quantum physics~\cite{ref:Purification}. Recently, a
state generation strategy based on the extraction of a state
through repeated measurements has been
proposed~\cite{ref:Nakazato_PRL,ref:Nakazato_PRA}. Since in most
cases this procedure allows to extract a pure state from a mixed
one, it has been addressed as a `purification.' Moreover, on the
basis of this idea, many applications have been proposed: the
extraction of
entanglement~\cite{ref:Nakazato_PRA,ref:Lidar_PRA-Paternostro} and
the initialization of multiple qubits would be useful for quantum
computation~\cite{ref:Nakazato_PRA,ref:qdot}; extensions of the
scheme enable us to establish entanglement between two spatially
separated systems via repeated measurements on an entanglement
mediator~\cite{ref:qpfe-separated}; in single trapped ions, the
extraction of angular-momentum Schr\"odinger-cat states has been
proposed~\cite{ref:Militello_PRA04} and the possibility of
steering the extraction of pure states through quantum Zeno effect
has been predicted~\cite{ref:Militello_PRA05}. The effect of the
environment during the process has been deeply
analyzed~\cite{ref:Militello_PRA07}. The scheme for the extraction
of quantum states is based on the following idea. When a quantum
system is put in interaction with a periodically measured one, the
initial state of the former system, $\rho(0)$, turns out to be
mapped into the state
$\aleph_k\hat{V}^k(\tau)\rho(0)[\hat{V}(\tau)^\dag]^k$, with
$\hat{V}(\tau)=\Bra{\phi_0}e^{-i\hat{H}\tau}\Ket{\phi_0}$, where
$\hat{H}$ is the Hamiltonian ($\hbar=1$) of the whole system (i.e.
the two interacting quantum systems one of which is repeatedly
measured), $\tau$ is the time between two measurements,
$\Ket{\phi_0}$ is the state of the subsystem projected by the
measurements, assuming that it is always the same, $k$ is the
number of observations, and $\aleph_k$ is the normalization
constant. The evolution described by the linear map
$\hat{V}(\tau)$ is a conditional one~\cite{ref:Nakazato_PRL}.
After a large number of measurements, the system is driven toward
a subspace which is given by the (right-) eigenspaces of
$\hat{V}(\tau)$ corresponding to the maximum (in modulus)
eigenavalues in its spectrum. If the extracted subspace is
one-dimensional, the final state (ideally reached after an
infinite number of measurements, in practical situations
approached after a finite and not too large number of steps) is
pure. A first expectation one can have about the behavior of
purity is that, starting from the initial value
$\trace[\rho(0)^2]$, it reaches the value $1$ {\em monotonically}.
In reality, as will become clear later, this is not always the
case. In fact, in many situations the purity of the system
oscillates and, passing through local minima and maxima, reaches
the final asymptotic value. This is an important point to be
understood since if one does not perform a sufficient number of
measurements the state which one gets is not only different from
the desired one, but also a state with lower purity than the
initial one.

In this paper we analyze the behavior of the purity of the state
of a two-level system interacting with a repeatedly measured one.
In the next section we derive an expression for the purity after
the $k$-th measurement and find suitable conditions that
characterize monotonic and non-monotonic behavior of this
quantity. In section \ref{sec:purity-specialsystem} we apply the
general formalism to a simple case, which could be of practical
interest. Finally, in section \ref{sec:lastsection} we give a
summary of the results.

\section{The purity after a finite number of measurements}\label{sec:purity-finite}

Let us consider a two-level system (\slvsys) interacting with
another system (\mstsys) which is repeatedly measured. Its state
after $k$ measurements of \mstsys\, is expressible as the $k$-th
power of the operator $\hat{V}(\tau)$, or simply $\hat{V}$,
applied to the initial state,
$\rho_k=\aleph_k\hat{V}^k(\tau)\rho(0)[\hat{V}^{\dag}(\tau)]^k$.
Assume that $\hat{V}$ is diagonalizable, i.e.
$\hat{V}=\lambda_1\Ket{u_1}\Bra{v_1}+\lambda_2\Ket{u_2}\Bra{v_2}$,
with $|\lambda_1| > |\lambda_2|$ the two eigenvalues, $\Bra{v_1}$
and $\Bra{v_2}$ the two left-eigenvectors, and $\Ket{u_1}$ and
$\Ket{u_2}$ the two independent (in general non-orthogonal)
normalized ($\BraKet{u_j}{u_j}=1$) right-eigenvectors. The
left-eigenvectors constitute a biorthonormal basis with the
right-eigenvectors, i.e. $\BraKet{v_i}{u_j}=\delta_{ij}$.
Therefore, the initial state $\rho_0=\rho(0)$ can be expanded as
\begin{subequations}
\begin{equation}\label{eq:expanded_initialstate}
\rho_0=a\Projector{u_1}+b\Projector{u_2}+c\KetBra{u_1}{u_2}+c^*\KetBra{u_2}{u_1}\,,
\end{equation}
with
\begin{align}
\label{eq:expanded_initialstate-eldef-a}
&a = \Bra{v_1}\rho_0\Ket{v_1}\ge 0\\
\label{eq:expanded_initialstate-eldef-b}
&b = \Bra{v_2}\rho_0\Ket{v_2}\ge 0\\
\label{eq:expanded_initialstate-eldef-c}
&c=\Bra{v_1}\rho_0\Ket{v_2}\, .
\end{align}
\end{subequations}
Its determinant is expressible as
\begin{equation}\label{eq:expanded_det_mappedstate}
\det\rho_0=(ab-|c|^2)|\BraKet{u_2}{u_1^\bot}|^2\,,
\end{equation}
with $\BraKet{u_1}{u_1^\bot}=0$ and
$\BraKet{u_1^\bot}{u_1^\bot}=1$. After $k$ steps, the initial
state is mapped into
\begin{subequations}
\begin{align}
\nonumber \rho_k=&%\aleph_k\,\hat{V}^k\,\rho_0\,\hat{V}^{\dag\,k}=
\nonumber \aleph_k\left[a|\lambda_1|^{2k}\Projector{u_1}+
b|\lambda_2|^{2k}\Projector{u_2}+\right.\\
&\left.c(\lambda_1\lambda_2^*)^k\KetBra{u_1}{u_2}+c^*(\lambda_1^*\lambda_2)^k\KetBra{u_2}{u_1}\right]\,,
\label{eq:expanded_mapped_initialstate}
\end{align}
with
\begin{equation}
\aleph_k^{-1}=a|\lambda_1|^{2k}+b|\lambda_2|^{2k}+\left(c(\lambda_1\lambda_2^*)^k\BraKet{u_2}{u_1}+c.c.\right)\,.
\end{equation}
\end{subequations}
The purity of this state is
\begin{subequations}
\begin{align}
\purity[\rho_k]:=\trace\rho_k^2=
1-\frac{2g^{2k}\,\det\rho_0}{(a+bg^{2k}+2\absc
g^k\cos\alpha_k)^2}\, , \label{eq:expanded_purity_mappedstate}
\end{align}
with
\begin{align}
&g=\left|\frac{\lambda_2}{\lambda_1}\right| < 1\,,\\
&\absc = |\,c \BraKet{u_2}{u_1}|\ge 0\,,\\
&\alpha_k=\arg\left[c\BraKet{u_2}{u_1}(\lambda_1\lambda_2^*)^k\right]\,.
\end{align}
\end{subequations}
If $g\not=0$, \eqref{eq:expanded_purity_mappedstate} gives also
the purity of the initial state for $k=0$. One immediately sees
that if the initial state is pure, $\purity[\rho_0]=1$
$\Leftrightarrow$ $\det\rho_0=0$, then the purity of the state
$\rho_k$ is always $1$. Instead, if the initial state is not a
pure one, oscillations of purity are possible, depending on the
parameters.

{\bf\em Local minima ---} In some cases, the action of the
operator $\hat{V}$ can diminish the purity of the state. To find
the relevant condition, let us impose
$\purity[\rho_1]<\purity[\rho_0]$, which immediately leads to
\begin{equation}\nonumber
\frac{g^2}{(a+bg^2+2g\absc\cos\alpha_1)^2}>\frac{1}{(a+b+2\absc\cos\alpha_0)^2}\,.
\end{equation}
Taking into account that the denominators are positive, one
eventually gets the condition
\begin{equation}\label{eq:purity_dimishingcondition}
a<bg-\frac{2g\absc}{1-g}(\cos\alpha_1-\cos\alpha_0)\, .
\end{equation}
Therefore, if \eqref{eq:purity_dimishingcondition} is satisfied
and the initial state is not pure, the purity of the state
decreases after the first measurement. Since the purity eventually
reaches the value $1$, the existence of a minimum of the purity is
guaranteed.

The relevant mechanism can be understood considering the very
special case where $\hat{V}$ is diagonalizable in the usual sense,
i.e. $\Ket{v_j}=\Ket{u_j}$, and the initial state is a mixture of
its two eigenstates, i.e.
$\rho(0)=\rho_{11}(0)\Projector{u_1}+\rho_{22}(0)\Projector{u_2}$,
so that $a=\rho_{11}(0)$ and $b=\rho_{22}(0)$. In such a case,
condition \eqref{eq:purity_dimishingcondition} just becomes
$\rho_{11}(0)<g\,\rho_{22}(0)$, which means that the population of
the state to be extracted should be smaller than that of the other
one times the parameter which determines how fast the process of
extraction is (indeed, remind that the smaller $g$ is, the smaller
the number of measurements required to extract the target state
is). In fact, since the state that will be extracted is
$\Ket{u_1}$, the extraction process lowers down the population of
$\Ket{u_2}$ and increases that of $\Ket{u_1}$. Depending on how
\lq slow\rq\ the process is (i.e., how many measurements are to be
performed to extract the target state), at some step the two
populations will reach closer values, which corresponds to a state
closer to the maximally mixed one than the initial state, and
therefore corresponds to a smaller value of the purity.

{\bf\em Local maxima --- } The two conditions
\begin{equation}
\purity[\rho_1]>\purity[\rho_0]\,, \qquad
\purity[\rho_1]>\purity[\rho_2]
\end{equation}
can be achieved simultaneously, determining the presence of a
local maximum at the first measurement. These two conditions are
compatible only if
\begin{equation}
b <
\frac{2\absc}{(1-g)^2(1+g)}\left[\cos\alpha_1-\cos\alpha_0+g(\cos\alpha_1-\cos\alpha_2)\right]
.
\end{equation}
The presence of local maxima is less intuitive to understand, even
though it can be forecasted mathematically. It is worth to note
that in the case wherein the operator $\hat{V}$ can be
diagonalized in the usual sense, local maxima are impossible,
since in such a case $\BraKet{u_2}{u_1}=0$ implies $\absc=0$ and
eventually $b<0$, which is incompatible with
\eqref{eq:expanded_initialstate-eldef-b}.

{\bf\em Condition for monotonic behavior --- } Here we give
general conditions to ensure that the purity increases toward the
target value, $1$. The condition expressing the monotonic behavior
of the purity from the $k_0$-th measurement on is
\begin{equation}
\purity[\rho_k]\ge \purity[\rho_{k-1}]\, , \qquad \forall k\ge
k_0\, ,
\end{equation}
which, after elementary manipulations, gives the necessary and
sufficient condition
\begin{equation}\label{eq:monotonic-condition}
b g^{2k-1} - \frac{2 g^k \absc
}{1-g}\left(\cos\alpha_k-\cos\alpha_{k-1}\right)-a \le 0\,, \qquad
\forall k\ge k_0\, .
\end{equation}
Since the absolute value of the second term on the left-hand side
is always not larger than $4 \absc g^k/(1-g)$ whatever $k\ge 1$
is, a sufficient condition is given by
\begin{equation}\label{eq:monotonic-suff-condition}
g^{2k} + \frac{4 g \absc}{(1-g)b}\,g^k - \frac{a\,g}{b} \le 0\, ,
\qquad \forall k\ge k_0\, ,
\end{equation}
which can be rewritten as
\begin{equation}\label{eq:monotonic-suff-condition-k}
G_- \le g^k \le G_+\,,\qquad k\ge k_0\, ,
\end{equation}
\begin{equation}\nonumber
G_\pm = -\frac{2g\absc}{(1-g)b}\pm\sqrt{\frac{4g^2\absc^2}{(1-g)^2
b^2} + \frac{a\,g}{b}}\, .
\end{equation}
Taking into account that $G_-\le 0$ and $0\le g < 1$, that implies
$g^{k+1} \le g^k$, \eqref{eq:monotonic-suff-condition-k} turns out
to be equivalent to
\begin{equation}\label{eq:monotonic-condition-k2}
g^{k_0} \le \sqrt{\frac{4g^2\absc^2}{(1-g)^2 b^2} +
\frac{a\,g}{b}}-\frac{2g\absc}{(1-g)b}\, ,
\end{equation}
which makes it easy to find the minimum number of measurements
sufficient to get monotonicity:
\begin{equation}\label{eq:monotonic-condition-k-final}
k\ge \log\left[\sqrt{\frac{4g^2\absc^2}{(1-g)^2 b^2} +
\frac{a\,g}{b}}-\frac{2g\absc}{(1-g)b}\right]/\log g\,.
\end{equation}
In other words, after a number of measurements not smaller than
the right-hand side of \eqref{eq:monotonic-condition-k-final} no
oscillation of purity is possible and the purity of the state of
the system monotonically reaches the asymptotic value $1$. When it
happens that $\absc\ll 0$ or $g\ll 1$, condition
\eqref{eq:monotonic-condition-k-final} reduces to
\begin{equation}\label{eq:simplified-condition}
k\ge \frac{1}{2} (1 + \log\frac{a}{b}\, / \log g)\, ,
\end{equation}
which is a simplified expression allowing to obtain qualitative
estimations of the number of measurements sufficient to have only
increasing purity. The same simplified condition comes out
directly and exactly from \eqref{eq:monotonic-condition} in the
special case $\lambda_1\lambda_2^*\in \mathbb{R}^+$, which implies
$\alpha_k=\alpha_{k-1}$, or if $\hat{V}$ can be diagonalized in
the usual sense, which implies $\absc=0$ whatever the initial
state is. In such cases, condition \eqref{eq:simplified-condition}
is equivalent to \eqref{eq:monotonic-condition}, and therefore
becomes a necessary and sufficient condition.

\section{A simple physical system}\label{sec:purity-specialsystem}

As a specific example, let us consider a system extensively
studied consisting of two qubits subjected to an interaction
preserving the number of excitations. A possible realization is
given by two two-level atoms subjected to a dipole-dipole
interaction. Assuming that the matrix elements of the dipole
operators are real, and neglecting the counter-rotating terms, one
reaches the following Hamiltonian (for details, see
Refs.~\cite{ref:Nakazato_PRA,ref:TwoSpinsInteracting}):
\begin{equation}
\hat{H}_\totsys
=\sum_{i=\slvsys,\,\mstsys}\frac{\Omega}{2}(1+\Sig{z}{i})
+\epsilon(\Sig{+}{\slvsys}\Sig{-}{\mstsys}
+\Sig{-}{\slvsys}\Sig{+}{\mstsys})
\end{equation}
where
$\Sig{z}{i}=\KetBraSub{\up}{\up}{i}-\KetBraSub{\dn}{\dn}{i}$,
$\Sig{+}{i}=\KetBraSub{\up}{\dn}{i}=(\Sig{-}{i})^\dag$, $\Omega$
is the Bohr frequency of the two-level system and $\epsilon$ is
the coupling constant.

The eigenstates of the Hamiltonian are the triplet and singlet
two-spin states:
\begin{subequations}
\begin{align}
\Ket{2}_\totsys &=\Ket{\up}_\slvsys\Ket{\up}_\mstsys,\displaybreak[0]\\
\Ket{1}_\totsys &=\frac{1}{\sqrt{2}}\bigl[
\Ket{\up}_\slvsys\Ket{\dn}_\mstsys+\Ket{\dn}_\slvsys\Ket{\up}_\mstsys
\bigr],\\
\Ket{0}_\totsys &=\Ket{\dn}_\slvsys\Ket{\dn}_\mstsys,\\
\Ket{s}_\totsys &=\frac{1}{\sqrt{2}}\bigl[
\Ket{\up}_\slvsys\Ket{\dn}_\mstsys-\Ket{\dn}_\slvsys\Ket{\up}_\mstsys
\bigr]\,.
\end{align}
\end{subequations}
The corresponding eigenenergies are $2\Omega$, $\Omega+\epsilon$,
$0$, and $\Omega-\epsilon$, respectively. In the case
$\Omega>\epsilon$, the state $\Ket{0}$ is the ground state. The
state $\Ket{s}$ is stable, not being coupled to any of the other
three states.

Repeatedly measuring the system \mstsys\, and finding it in the
state
$\Ket{\theta}_\mstsys=\cos\frac{\theta}{2}\Ket{\up}_\mstsys+\sin\frac{\theta}{2}\Ket{\dn}_\mstsys$
(the most general state should include a phase factor, here
assumed equal to zero for simplicity), the system \slvsys\, is
subjected to a non-unitary evolution governed by the
operator~\cite{ref:Nakazato_PRA}
\begin{align}
\nonumber\hat{V}=&\left(e^{-2i\Omega t}\cos^2\frac{\theta}{2}+e^{-i\Omega t}\cos\epsilon t\,\sin^2\frac{\theta}{2}\right)\KetBra{\up}{\up}\\
&\nonumber + \left(\sin^2\frac{\theta}{2}+e^{-i\Omega
t}\cos\epsilon t\,\cos^2\frac{\theta}{2}\right)\KetBra{\dn}{\dn}
\\
&-i e^{-i\Omega t}\sin\epsilon
t\sin\frac{\theta}{2}\cos\frac{\theta}{2}\left(\KetBra{\up}{\dn} +
\KetBra{\dn}{\up}\right)\, ,\label{eq:VOperator}
\end{align}
where we have omitted the subscript $\slvsys$ of the states.

\begin{figure}
  \begin{center}
    \includegraphics[width=0.46\textwidth, height=0.3\textwidth]{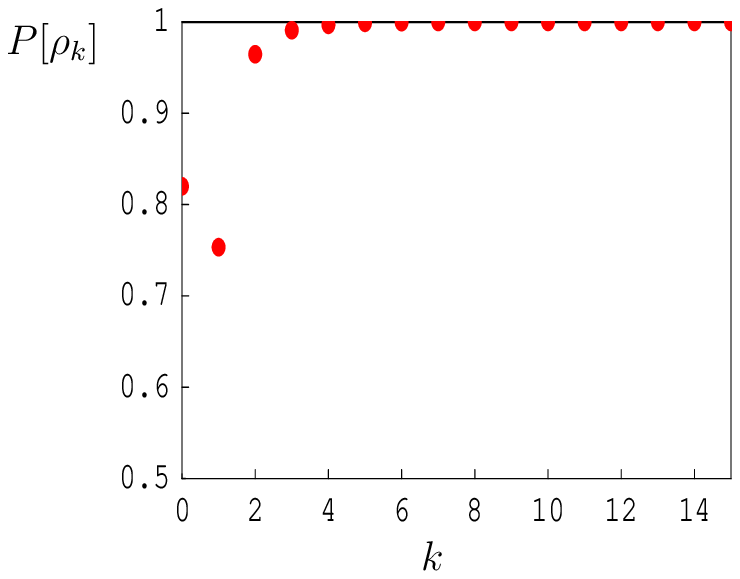}
  \end{center}
  \caption{The purity as a function of the number of measurements,
  starting with $\rho(0)=0.1\,\Projector{\dn}+0.9\,\Projector{\up}$,
  in the parameter region characterized by $\epsilon\tau=7.82$,
  $\theta=2.25$, $\epsilon/\Omega=0.1$.} \label{fig:RelativeMinimum}
\end{figure}

\begin{figure}
  \begin{center}
    \includegraphics[width=0.46\textwidth, height=0.3\textwidth]{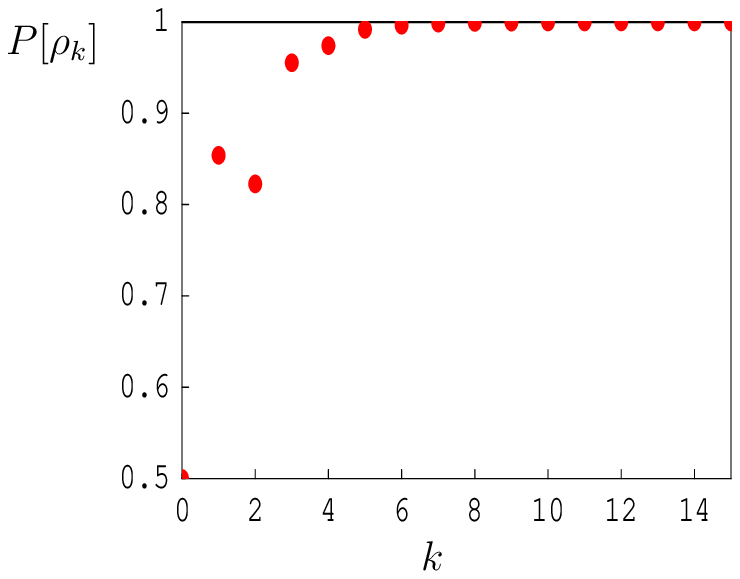}
  \end{center}
  \caption{The purity as a function of the number of measurements,
  starting with  $\rho(0)=0.5\,\Projector{\dn}+0.5\,\Projector{\up}$,
  in the parameter region characterized by $\epsilon\tau=2.50$, $\theta=1.0$, $\epsilon/\Omega=0.1$.} \label{fig:RelativeMaximum}
\end{figure}

This very simple physical situation exhibits non-monotonic
behavior associated with the presence of minima and maxima of the
purity. Consider the following situations described in the
figures. In figure \ref{fig:RelativeMinimum} we see that starting
from $\rho(0)=0.1\,\Projector{\dn}+0.9\,\Projector{\up}$,
repeatedly measuring the state of \mstsys\, characterized by
$\theta = 2.25$ with time interval $\tau=7.82 / \epsilon$, a
minimum of the purity occurs at the first measurement. Instead,
figure \ref{fig:RelativeMaximum} shows that starting from the
maximally mixed state and measuring the state of the system
\mstsys\, characterized by $\theta=1.0$ with time interval
$\tau=2.50 / \epsilon$, the purity exhibits a local maximum at the
first measurement and a local minimum at the second measurement.
From the point of view of applications, it could be important to
give the conditions under which non-monotonic behavior of purity
is avoided.

{\bf\em Measuring the state $\Ket{\up}_\mstsys$ --- } As a very
special case, assume that the system \mstsys\, is repeatedly
measured (after each $\tau$ such that $\epsilon\tau\not=q\pi$,
with $q\in\mathbb{Z}$, to avoid the case $g=1$) and found in the
up-state $\Ket{\up}_{\mstsys}$, so that
\begin{equation}
\hat{V}=e^{-2i\Omega t}\KetBra{\up}{\up}+ e^{-i\Omega
t}\cos\epsilon t\,\KetBra{\dn}{\dn}\, ,
\label{eq:VOperator-special-up}
\end{equation}
in correspondence to which we get $g=|\cos\epsilon t|$, while the
eigenvectors are $\Ket{u_1}=\Ket{v_1}=\Ket{\up}$ and
$\Ket{u_2}=\Ket{v_2}=\Ket{\dn}$. In such a situation, the
coefficients $a$, $b$ and $c$ are the usual matrix elements of the
density operator $\rho(0)$ with respect to an orthonormal basis.
Starting from the general state
$\rho(0)=\probup\Projector{\up}+(1-\probup)\Projector{\dn}+c\KetBra{\up}{\dn}+c^*\KetBra{\dn}{\up}$,
we have $a=\probup$, $b=1-\probup$ and $\absc=0$, so that the
(necessary and sufficient) condition for monotonicity is
\begin{equation}\label{eq:threshold-measup}
k \ge \frac{1}{2} (1 + \log\frac{\probup}{1-\probup}\, /
\log|\cos\epsilon\tau| )\, .
\end{equation}
If $\probup>0.5$, the right-hand side is smaller than unity and
only monotonic increase of the purity is possible. On the
contrary, if $\probup$ is small enough to have
$\log(\probup/(1-\probup)) < \log|\cos\epsilon\tau|$,
non-monotonic behavior is possible.

\begin{figure}
  \begin{center}
    \includegraphics[width=0.46\textwidth, height=0.32\textwidth]{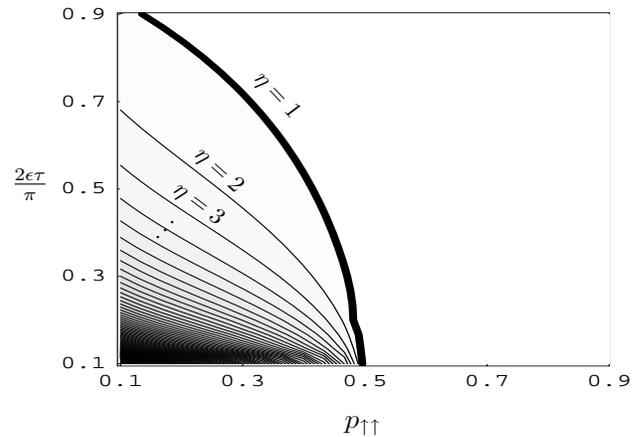}
  \end{center}
  \caption{Contour plot of the quantity $\eta=\max\{0,\frac{1}{2} (1 + \log(\probup/(1-\probup))\, /
  \log|\cos\epsilon\tau|)\}$, vs the population of $\Ket{\up}$, $\probup$, and the dimensionless time between two
  measurements, $2\epsilon\tau / \pi$, both in the range $[0.1,\,0.9]$.
  The bold line corresponds to $\eta=1$, which discriminates between
  monotonic behavior of the purity ($\eta < 1$ on the right-hand side of the line) and oscillations ($\eta > 1$ in the left-hand side).
  In the dark region (in the bottom of the left-hand side) the value of $\eta$ is very high, meaning that one needs a large number of measurements to guarantee monotonicity.}
  \label{fig:Threshold}
\end{figure}

Figure \ref{fig:Threshold} shows the threshold for the number of
measurements necessary to guarantee monotonicity as a function of
the population of the target state in the initial condition
($\probup$) and of the time $\tau$. Since the right-hand side of
\eqref{eq:threshold-measup} can become negative (in which case we
get monotonic behavior, as well as for any value smaller than
unity), we consider its nonnegative counterpart, i.e.
$\eta=\max\{0,\frac{1}{2} (1 + \log(\probup/(1-\probup))\,/
\log|\cos\epsilon\tau|)\}$, without loss of information. It is
well visible that for most of the values of the parameters (on the
right-hand side of the figure, beyond the curve $\eta=1$) the
threshold is lower than unity, meaning that only monotonic
increase of the purity occurs. Instead, for low values of
$\probup$ and $\tau$ the threshold becomes higher than unity,
meaning that non-monotonic behavior occurs.

If the state $\Ket{\dn}_\mstsys$ is measured we get essentially
the same results, even the same expression of the threshold,
provided the replacement $\probup \rightarrow 1-\probup$ has been
done.

\section{Summary}\label{sec:lastsection}

Summing up, we have analyzed the problem of non-monotonic behavior
of the purity of the state of a physical system when it is
subjected to a process of extraction of pure states by repeatedly
measuring a quantum system interacting with the former one. The
knowledge of this phenomenon is of fundamental importance, since
it brings to the light interesting features of the transient from
the initial state to the final one. Moreover, in the applications
it is crucial to determine the minimum number of steps necessary
to get the monotonic increase of purity. Otherwise, the
purification process could be counterproductive, resulting in a
diminishing of the purity. The general results we have found for a
two-level system, reported in section \ref{sec:purity-finite} and
then applied to a specific physical system in section
\ref{sec:purity-specialsystem}, allow to overcome this problem and
could be of interest in practical applications of this scheme for
the generation of pure states.


\begin{thebibliography}{99}

\bibitem{ref:q_tech}
For reviews on quantum information, see: \textit{The Physics of
Quantum Information}, edited by D. Bouwmeester, A. Zeilinger, and
A. Ekert (Springer-Verlag, Berlin, 2000); M.~A. Nielsen and I.~L.
Chuang, \textit{Quantum Computation and Quantum Information}
(Cambridge University Press, Cambridge, 2000); C.~H. Bennett and
D.~P. DiVincenzo, \textit{Nature} (London) \textbf{404}, 247
(2000); A. Galindo and M.~A. Mart\'{\i}n-Delgado, \textit{Rev.
Mod. Phys.} \textbf{74}, 347 (2002); P. Kok, W. J. Munro, K.
Nemoto, T. C. Ralph, J. P. Dowling, and G. J. Milburn,
\textit{Rev. Mod. Phys.} \textbf{79}, 135 (2007).

\bibitem{ref:Purification}
C.~H. Bennett, G. Brassard, S. Popescu, B. Schumacher, J.~A.
Smolin, and W.~K. Wootters, \textit{Phys. Rev. Lett.} \textbf{76},
722 (1996); \textbf{78}, 2031(E) (1997); C.~H. Bennett, D.~P.
DiVincenzo, J.~A. Smolin, and W.~K. Wootters, \textit{Phys. Rev.}
A \textbf{54}, 3824 (1996); J.~I. Cirac, A.~K. Ekert, and C.
Macchiavello, \textit{Phys. Rev. Lett.} \textbf{82}, 4344 (1999);
T. Yamamoto, M. Koashi, \c{S}. K. {\"O}zdemir, and N. Imoto,
\textit{Nature} (London) \textbf{421}, 343 (2003); Z. Zhao, T.
Yang, Y.-A. Chen, A.-N. Zhang, and J.-W. Pan, \textit{Phys. Rev.
Lett.} \textbf{90}, 207901 (2003); J.-W. Pan, S. Gasparoni, R.
Ursin, G. Weihs, and A. Zeilinger, \textit{Nature} (London)
\textbf{423}, 417 (2003); M. Ricci, F. De Martini, N. J. Cerf, R.
Filip, J. Fiur\'{a}\v{s}ek, and C. Macchiavello, \textit{Phys.
Rev. Lett.} \textbf{93}, 170501 (2004); A. Franzen, B. Hage, J.
DiGuglielmo, J. Fiur\'{a}\v{s}ek, and R. Schnabel, \textit{Phys.
Rev. Lett.} \textbf{97}, 150505 (2006).

\bibitem{ref:Nakazato_PRL}
H. Nakazato, T. Takazawa, and K. Yuasa, {\it Phys.\ Rev.\ Lett.}
{\bf 90}, 060401 (2003).

\bibitem{ref:Nakazato_PRA}
H. Nakazato, M. Unoki, and K. Yuasa, \textit{Phys. Rev.} A
\textbf{70},  012303 (2004).

\bibitem{ref:Lidar_PRA-Paternostro}
L.-A. Wu, D.~A. Lidar, and S. Schneider, \textit{Phys. Rev.} A
\textbf{70}, 032322 (2004); M. Paternostro and M.~S. Kim,
\textit{New J. Phys.} \textbf{7}, 43 (2005).

\bibitem{ref:qdot}
K. Yoh, K. Yuasa, and H. Nakazato, \textit{Physica} E \textbf{29},
674 (2005); K. Yuasa, K. Okano, H. Nakazato, S. Kashiwada, and K.
Yoh, \textit{AIP Conf. Proc.} \textbf{893}, 1109 (2007).

\bibitem{ref:qpfe-separated}
G. Compagno, A. Messina, H. Nakazato, A. Napoli, M. Unoki, and K.
Yuasa, \textit{Phys. Rev.} A \textbf{70}, 052316 (2004); K. Yuasa
and H. Nakazato, \textit{Prog. Theor. Phys.} \textbf{114}, 523
(2005); \textit{J. Phys.} A \textbf{40}, 297 (2007).

\bibitem{ref:Militello_PRA04}
B. Militello and A. Messina, \textit{Phys. Rev.} A \textbf{70},
033408 (2004); \textit{Acta Phys. Hung.} B \textbf{20}, 253
(2004).

\bibitem{ref:Militello_PRA05}
B. Militello, H. Nakazato, and A. Messina, \textit{Phys. Rev.} A
\textbf{71}, 032102 (2005); B. Militello, A. Messina, and H.
Nakazato, \textit{Opt. Spectrosc.} \textbf{99}, 438 (2005).

\bibitem{ref:Militello_PRA07}
B. Militello, K. Yuasa, H. Nakazato, and A. Messina, \textit{Phys.
Rev.} A \textbf{76}, 042110 (2007); H. Nakazato, K. Yuasa, B.
Militello, and A. Messina, {\it Estimation of the
repeatedly-projected reduced density matrix under decoherence},
ArXiv: (quant-ph) 0711.2751.

\bibitem{ref:TwoSpinsInteracting}
S. Nicolosi, A. Napoli, A. Messina, and F. Petruccione,
\textit{Phys. Rev.} A \textbf{70},  022511  (2004).

\end{thebibliography}
\end{document}